\documentclass{rbfin}

\begin{document}
\selectlanguage{english}
\frenchspacing


\title{Analysis on the Influence of Synchronization Error on Fixed-filter Active Noise Control } 
\otitle{} 

\shorttitle{Brazilian Review of Finance: A template} 

\nota{Submitted on February 8, 2023.
Revised on May 25, 2023.
Accepted on June 9, 2023.
Published online in June 2023.
Editor in charge: Mr Editor.
}

\author{
\bf{Guo Yu}%
\footnote{Author One affiliation, Brazil: \email{E220059@e.ntu.edu.sg}}
}
\autor{Author One et al., 2023} 

\maketitle
\pagina{1}
\rbfd{
\href
{http://bibliotecadigital.fgv.br/ojs/index.php/rbfin/index}
{Brazilian Review of Finance (Online)},
Rio de Janeiro,
\textcolor{BrickRed}{Vol. XX,}
\textcolor{BrickRed}{No. Y,}
\textcolor{BrickRed}{August 2023,}
\textcolor{BrickRed}{pp. x--xx}
\qquad ISSN 1679-0731, ISSN online 1984-5146}
\rbfc{\copyright
2023
\href
{https://www.sbfin.org.br}
{Sociedade Brasileira de Finanças},
under a
\href
{http://creativecommons.org/licenses/by/3.0}
{Creative Commons Attribution 3.0 license} 
}
\rbfe{
\href
{http://bibliotecadigital.fgv.br/ojs/index.php/rbfin/index}
{Brazilian Review of Finance (Online)}
XX(Y),
2023
}
\noindent

\begin{abstract}
\foreignlanguage{english}{%
\textbf{Abstract}\ 
The efficacy of active noise control technology in mitigating urban noise, particularly in relation to low-frequency components, has been well-established. In the realm of traditional academic research, adaptive algorithms, such as the filtered reference least mean square method, are extensively employed to achieve real-time noise reduction in many applications. Nevertheless, the utilization of this technology in commercial goods is often hindered by its significant computing complexity and inherent instability. In this particular scenario, the adoption of the fixed-filter strategy emerges as a viable alternative for addressing these challenges, albeit with a potential trade-off in terms of noise reduction efficacy. This work aims to conduct a theoretical investigation into the synchronization error of the digital Active Noise Control (ANC) system.      

}
\end{abstract}


\section{Problem statement}\label{sec-intro}
Active Noise Control (ANC), also known as active noise cancellation or active noise reduction, is a technology and technique used to reduce or eliminate unwanted ambient noise by generating sound waves that are precisely phase-inverted to the incoming noise~\cite{shi2023active,hansen2002understanding,elliott1993active,kuo1999active}. The goal of ANC is to create a quieter and more comfortable environment for individuals in various settings, such as homes~\cite{lam2020active,lam2020active1,shi2023computation,hasegawa2019multi,he2019exploiting,hasegawa2018window,lam2018active,shi2017algorithms,lam2023anti}, offices, vehicles, and even headphones~\cite{shen2023implementations,shen2022multi,shen2022multiA,shen2022hybrid,shen2022adaptive,shen2021implementation,shen2021wireless,shen2021alternative}. ANC works on the principle of superposition of sound waves. When two sound waves with equal amplitude but opposite phases (180-degree phase difference) are combined, they interfere destructively, effectively canceling each other out. This means that if you can create a sound wave that is exactly out of phase with the unwanted noise, they will cancel each other, resulting in reduced or eliminated noise. ANC systems are often adaptive, meaning they continuously monitor the incoming noise and adjust the anti-noise signal in real-time. This adaptation allows the system to respond to changes in the noise environment, making it effective in various conditions. Nevertheless, the utilization of adaptive algorithms, such as those based on the least mean squares (LMS) methodology, often raises stability concerns. Additionally, the sluggish convergence rate associated with these algorithms can significantly impact their effectiveness in effectively handling dynamic noise~\cite{wen2020convergence,lam2021ten,shi2021fast,shi2020active}. 

In the practical applications, such as in commercial ANC headphones, the processor initially performs offline estimation of the secondary paths. Subsequently, the control filters are trained using an adaptive filter technique. Once the optimal control filters have been obtained, the processor will cease training and proceed to utilize these filters for the purpose of filtering the reference signal~\cite{shi2020feedforward,shi2018novel,wen2020using}. By implementing the aforementioned technique, the ANC system is able to provide effective noise suppression while solving the system stability issue caused by the adaptive active control algorithms~\cite{shi2017effect,shi2021optimal,shi2019two,shi2019optimal,shi2021optimal3,shi2023frequency,lai2023mov}. The active control technique referred to as the fixed-filter approach has been extensively employed in commercial applications and has garnered significant attention in academic research~\cite{shi2022selective,luo2023deep,shi2023transferable,luo2023performance,luo2022hybrid}.  

The rationale behind the effectiveness of the aforementioned strategy is readily apparent. The coefficients of optimal filters remain unchanged when the physical environment and primary noise remain fixed. In a practical context, the majority of ANC applications fulfill these two criteria. It can be argued that in a given physical environment, many distinct ANC systems with identical Analog-to-Digital Converter (ADC), Digital-to-Analog Converter (DAC), and sampling rates will possess identical optimal control filter coefficients to effectively address a shared primary noise source. By extrapolating this assumption, it is possible to utilize a high-cost yet high-performance processor for training the best control filter. Subsequently, the obtained coefficients can be implemented on a less expensive but lower-performance processor to effectively mitigate the primary noise in the identical physical setting. The less expensive processor just requires the completion of fixed filters, specifically the Finite Impulse Response (FIR) with the ideal control filter's coefficients. This approach eliminates the requirement for adaptive filters, resulting in a substantial reduction in computational workload. 
\begin{figure}[!t]
    \centering
    \includegraphics[width=8.5cm]{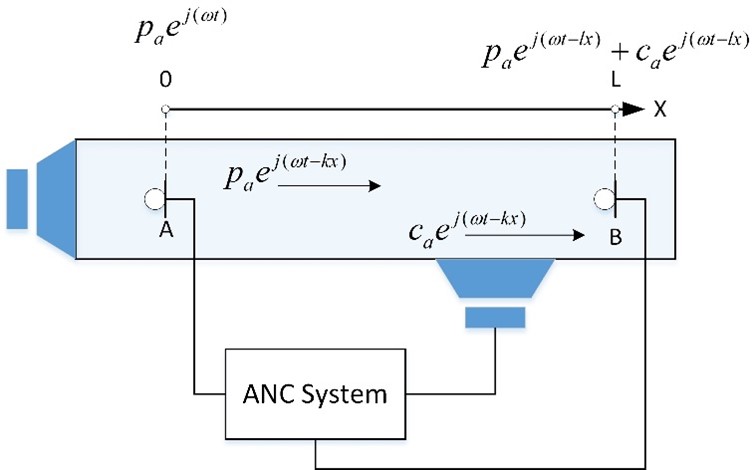}\caption{The single-channel feedforward ANC system~\cite{shi2020algorithms}.}
    \label{fig:enter-label}
\end{figure}

Nevertheless, the empirical findings of the experiment diverge from this premise. Regardless of the type of primary noise employed or the specific physical environment, it is evident that the noise reduction achieved by a fixed filter is consistently inferior than that of the comparable adaptive Active Noise Control (ANC) filter. This is due to the oversight of the fact that two digital hardware configurations, even if they are identical systems that are powered on at different times, must possess distinct system clocks. The disparity can be attributed to two distinct components: variations in frequency and disparities in the early phase of the system clock. The utilization of several system clocks will result in distinct sample clocks for the analog-to-digital converter (ADC), thereby significantly impacting the performance of the fixed filter. In the subsequent study, we designate the sample clock of the adaptive ANC filter system, which obtains the optimal control filters, as the reference clock. We refer to the disparity in initial phase and frequency between the sampling clocks of the fixed filter and adaptive ANC filter as the initial phase error and frequency error, respectively.

\section{The influence of frequency error on fixed filter}

To simplify the analysis, we assume the primary noise is a plane wave that transmits along the x-axis in a free field, and the ANC system is a single-channel feedforward structure. The reference microphone is at location A, the original point of the x-axis. The error microphone is located at the B of the axis, whose distance is l. The pressure of the primary noise is
\begin{equation}
    p(t,x)=p_a e^{j(\omega_0t-kx)},
\end{equation}
In this context, $\omega_0$ represents the frequency of the primary noise, $p_a$ denotes the amplitude of the plane wave, and $k$ is defined as $\frac{\omega_0}{c_0}$, where $c_0$ represents the speed of sound in the air. The pressure of the primary noise at the reference microphone's location is denoted as $p_a e^{j(\omega_0 t)}$, whereas at the error microphone's location, it is represented as $p_a e^{j(\omega_0 t-kl)}$, which is referred to as the disturbance noise. 
\begin{figure}
    \centering
    \includegraphics[width=8.5cm]{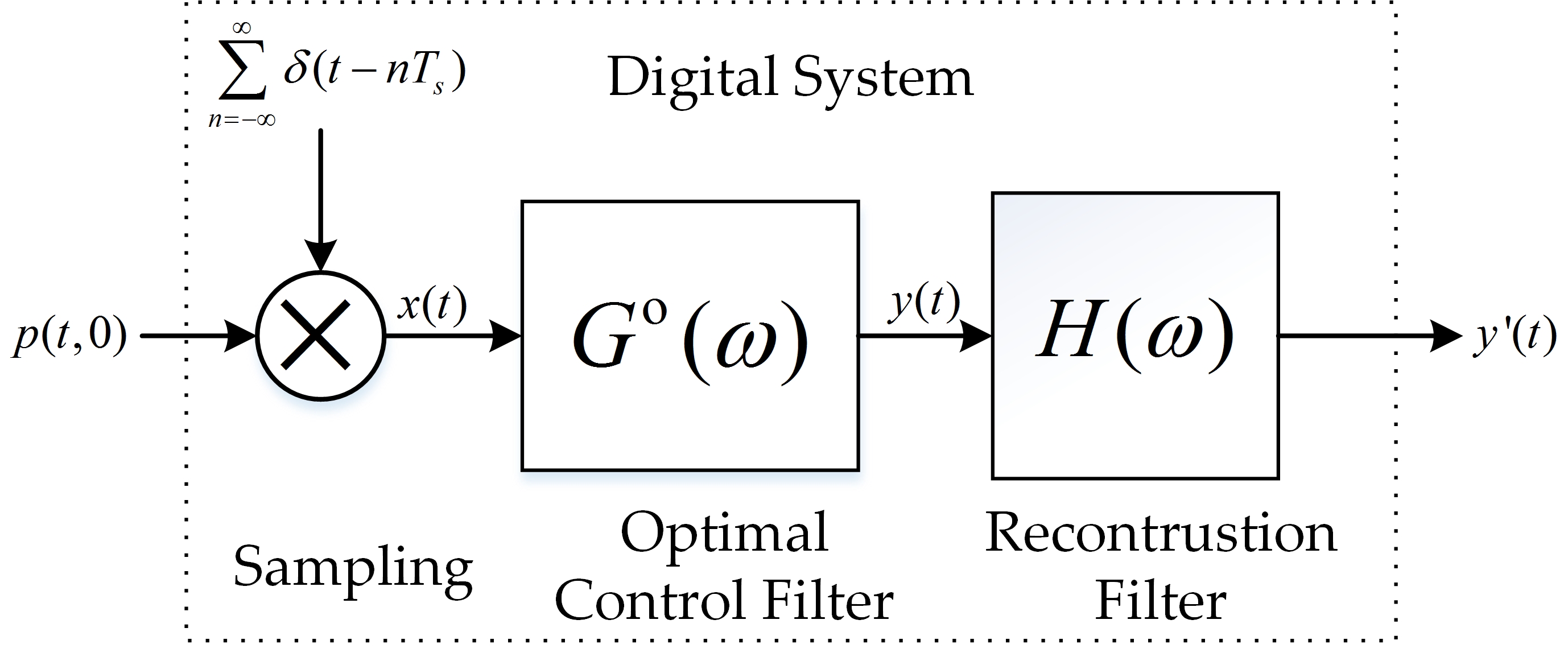}
    \caption{Block diagram of the ANC system on processor}
    \label{fig:2}
\end{figure}

When the processor gains the optimal control filter, it stops the adaptive process, and the block diagram of the ANC system on it can be expressed as Figure 2. The sample period of the Analog-Digital convertor is $T_s$, so the reference signal should be
\begin{equation} \label{2)} 
\begin{array}{l} {x(t)=p(t,0)|_{t=nT_{s} } =p(t,0)\sum _{n=-\infty }^{\infty }\delta (t-nT_{s} ) } \\ {\mathrm{\; \; \; \; \; \; \; }=p_{a} \sum _{n=-\infty }^{\infty }e^{j\omega _{0} t} \delta (t-nT_{s} ) } \end{array}.
\end{equation} 
Its Fourier transform is given by
\begin{equation} \label{3)} 
X(\omega )=\frac{2\pi p_{a} }{T_{s} } \sum _{j=-\infty }^{+\infty }\delta (\omega -\omega _{0} -j\frac{2\pi }{T_{s} } )
\end{equation} 
The impulse response of the optimal control filter can be expressed as 
\begin{equation} \label{4)} 
g^{o} (t)=\sum _{n=0}^{N-1}w_{n} \delta (t-nT_{s} ).  
\end{equation} 
The length of the optimal control filter is set to $N$, and its frequency response is obtained from  
\begin{equation} \label{5)} 
G^{o} (\omega )=\sum _{n=0}^{N-1}w_{n} e^{-j\omega nT_{s} }.  
\end{equation} 
To reconstruct the digital signal to the analog signal, we build a reconstruction filter whose frequency response is obtained from
\begin{equation} \label{6)} 
H(\omega )=\left\{\begin{array}{c} {T_{s} \mathrm{\; \; \; }0\le \omega <\frac{\pi }{T_{s} } } \\ {0\mathrm{\; \; \; }\text{others}} \end{array}\right. 
\end{equation} 

As mentioned in the previous assumption, the sound transmits in a free field and doesn't have any transmission attenuation. Therefore, the secondary path only brings some delays to the primary noise. The transfer function of the secondary path can be described as 
\begin{equation} \label{7)} 
S(\omega )=e^{-j\omega t_{s} }.  
\end{equation} 
The anti-noise at the location of the error microphone should be 
\begin{equation} \label{8)} 
c(t,l)=p(t,0)*g(t)*h(t)*s(t){\rm =}p_{a} \sum _{n=0}^{N-1}w_{n} e^{j\omega _{0} (t-nT_{s} {\rm -}t_{s} )}.   
\end{equation} 
The anti-noise wave can completely cancel the disturbance noise at the location of the error microphone. So,
\begin{equation} \label{9)} 
p(t,l)+c(t,l)=0.   
\end{equation} 
By substituting \eqref{8)} to \eqref{9)}, we can derive
\begin{equation} \label{10)} 
p(t,l)=-p_{a} \sum _{n=0}^{N-1}w_{n} e^{j\omega _{0} (t-nT_{s} {\rm -}t_{s} )}.   
\end{equation} 
\begin{figure}[!t]
    \centering
    \includegraphics[width=9.5cm]{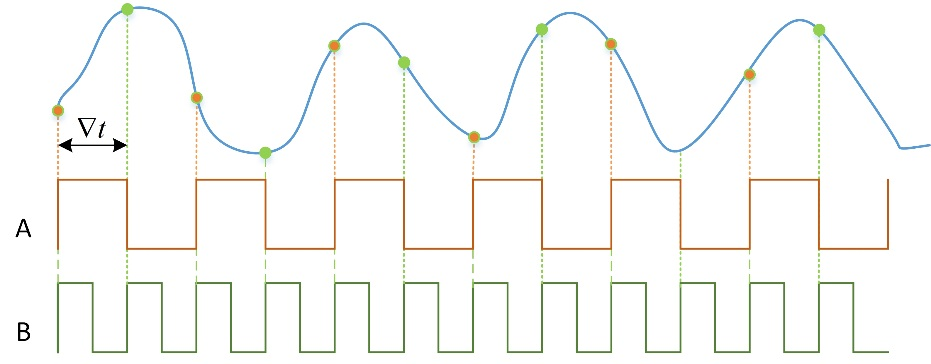}
    \caption{Reference sampling clock vs error sampling clock}
    \label{fig:3}
\end{figure}

When we reload the coefficients of the control filter into new digital hardware, the system clocks of these systems probably have a slight difference in frequency, even if they use the frequency sources with the same rate frequency output. The different system clocks must trigger them to own different sampling clocks. As Figure 3 shows, A denotes the reference sampling clock of the digital system shown in Figure 2, and B represents the error sampling clock of another system with the same control filter and hardware as the previous one but a different system clock. $\mathrm{\nabla }t$ is the difference between the reference and error sampling clock. The primary noise sampled by the error sampling clock will bring in some error. Then, the new reference signal can be written as
\begin{equation} \label{11)} 
    \begin{split}
        x'(t)&=p(t,0)|_{t=n(T_{s} +\nabla t)} =x_{a} (t)\sum _{n=-\infty }^{\infty }\delta (t-nT_{s} -n\nabla t) \\
        & =p_{a} \sum_{n=-\infty }^{\infty }e^{j\omega _{0} t} \delta (t-nT_{s} -n\nabla t).
    \end{split}
\end{equation} 
Its Fourier transform is given by 
\begin{equation} \label{12)} 
X^{'} (\omega )=\frac{2\pi p_{a} }{T_{s} +\nabla t} \sum _{j=-\infty }^{+\infty }\delta (\omega -\omega _{0} -j\frac{2\pi }{T_{s} +\nabla t} ). 
\end{equation}
Due to the optimal control filter using the same error sampling clock, its impulse response will also change to be
\begin{equation} \label{13)} 
g^{o} {}^{'} (n)=\sum _{n=0}^{N-1}w_{n} \delta (t-nT_{s} -n\nabla t),  
\end{equation} 
whose frequency response is given by 
\begin{equation} \label{14)} 
G^{o} {}^{'} (\omega )=\sum _{n=0}^{N-1}w_{n} e^{-j\omega n(T_{s} {\rm +}\nabla t)}.   
\end{equation} 
To reconstruct the digital signal to its analog signal, we build the reconstruction filter as 
\begin{equation} \label{15)} 
H^\prime (\omega )=\left\{\begin{array}{c} {T_{s} +\nabla t\mathrm{\; \; \; \; \; \; }0\le \omega <\frac{\pi }{T_{s} +\nabla t} } \\ {0\mathrm{\; \; \; \; \; \; \; \; \; \; \; \; \; \; }\text{others}} \end{array}\right.  
\end{equation} 

Hence, the new anti-noise is obtained from 
\begin{equation} \label{16)} 
c'(t,l)=x'(t)*g^{o} {}^{'} (t)*h(t)*s(t)=p_{a} \sum _{n=0}^{N-1}w_{n} e^{j\omega _{0} [t-n(T_{s} +\nabla t){\rm -}t_{s} ]},   
\end{equation} 
and the pressure on the error microphone is stated to be
\begin{equation} \label{17)} 
e(t)=p(t,l)+c'(t,l)=w_{1} p_{a} e^{j\omega _{0} (t-T_{s} )} (e^{j\omega _{0} \nabla t} -1).  
\end{equation} 
The magnitude of the error signal is represented as 
\begin{equation} \label{18)} 
||e(t)||_{2} =2w_{1}^{2} p_{a}^{2} [1-\cos (\omega _{0} \nabla t)].   
\end{equation} 

Based on the Nyquist principle  
\begin{equation} \label{19)} 
\left\{\begin{array}{c} {\frac{1}{T_{s} } >\frac{\omega _{0} }{\pi } } \\ {\frac{1}{T_{s} +\nabla t} >\frac{\omega _{0} }{\pi } } \end{array}\right.  
\end{equation} 
we can figure out $-\pi <{\omega }_0\mathrm{\nabla }t<\pi $. From the equation \eqref{18)} and \eqref{19)}, we can know when $\mathrm{\nabla }\mathrm{t}$ equals to 0, the anti-noise can counterbalance to the effect of disturbance noise. But $\mathrm{|}\mathrm{\nabla }\mathrm{t|}$ increase, the pressure of residual sound at the error microphone will also increase.

\section{The influence of initial phase error }

With the advancement of new technologies, numerous frequency sources with high output frequency accuracy have emerged. These sources are extensively utilized in high-precision digital systems. An exemplary illustration is the atomic oscillator. A collective of atomic oscillators has the ability to generate a frequency that is the same across all members. The underlying premise is straightforward: Given the uniformity of atoms inside a given element, their absorption or emission of energy should result in similar frequencies. Hence, in the event that both digital systems employ the same atomic oscillator, they have the potential to obtain an equivalent frequency. 
\begin{figure}[!t]
    \centering
    \includegraphics[width=9.5cm]{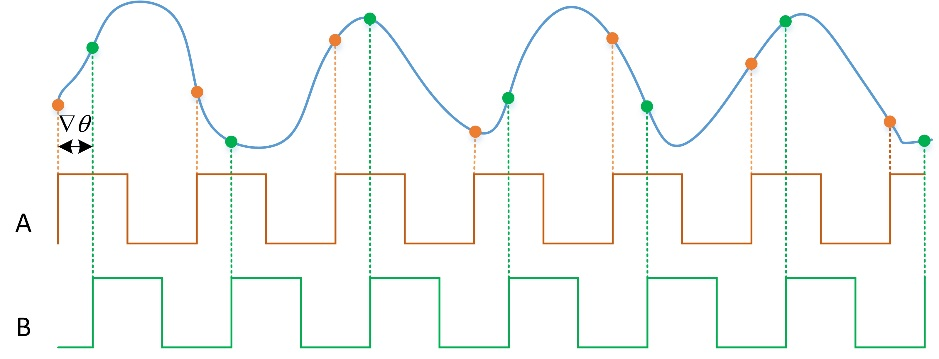}
    \caption{Reference sampling clock and error sampling clock with different initial phase}
    \label{fig:4}
\end{figure}

Nevertheless, it is essential to acknowledge that a digital system may exhibit varying initial phases of system clocks depending on the moment it is powered. The occurrence of an initial phase error in the initial phase will also result in a comparable initial phase error in the sampling clock. The phenomena are illustrated in Figure 4. The depicted image shows the reference sampling clock, denoted by A, and the error sampling clock, represented by B. The A and B sampling clocks exhibit identical frequencies but with distinct beginning phases. The symbol $\mathrm{\nabla }\theta$ represents the initial phase difference, referred to as the initial phase error. 

\subsection{Narrow Band Primary Noise }

In general situations, $\mathrm{\nabla }\theta$ the same sampling clock causes triggered at different times. In our application, $\mathrm{\nabla }\theta $ comes from the different beginnings of adaptive filter sampling clock and fixed filter sampling clock. Based on experience, this initial phase error should belong to a uniform disturbance. Its probability density function is
\begin{equation} \label{20)} 
P(\nabla \theta )=\left\{\begin{array}{c} {\frac{1}{2\pi } \mathrm{\; \; \; }\nabla \theta \in \left[0,2\pi \right]} \\ {0\mathrm{\; \; \; }\nabla \theta \notin \left[0,2\pi \right]} \end{array}\right.  
\end{equation} 
Because of the affection of this initial phase error, the sampled reference signal will also change to 
\begin{equation} \label{21)} 
\begin{split}
    x''(t)&=p(t,0)|_{t=nT_{s} +\frac{\nabla \theta }{2\pi } T_{s} } =x_{a} (t)\sum _{n=-\infty }^{\infty }\delta (t-nT_{s} -\frac{\nabla \theta }{2\pi } T_{s} ) \\
    &=p_{a} \sum _{n=-\infty }^{\infty }e^{j\omega _{0} t} \delta (t-nT_{s} -\frac{\nabla \theta }{2\pi } T_{s} ).
\end{split}
\end{equation} 
From the above equation, we can figure out that the initial phase error brings a constant delay for sampling impulses. The equation \eqref{21)} can also be written as
\begin{equation} \label{22)} 
x''(t)=p_{a} \sum _{n=-\infty }^{\infty }e^{j\omega _{0} (t+\frac{\nabla \theta }{2\pi } T_{s} )} \delta (t-nT_{s} ).  
\end{equation} 
In this equation, the initial phase error just makes the analog reference signal have a time shift.  Therefore, the implementation of anti-noise technology should be considered to be 
\begin{equation} \label{23)} 
c''(t,l)=x''(t)*g(t)*h(t)*s(t)=p_{a} \sum _{n=0}^{N-1}w_{n} e^{j\omega _{0} (t-nT_{s} +\frac{\nabla \theta }{2\pi } T_{s} {\rm -}t_{s} )}.   
\end{equation}
Therefore, the error (we call it residual error) brought by the initial phase error at noise reduction can be expressed as 
\begin{equation} \label{24)} 
e(\nabla \theta )=E\{ [p(t,l)+c''(t,l)]^{2} \} -E\{ [p(t,l)+c(t,l)]^{2} \} . 
\end{equation} 
The equation can rewritten as 
\begin{equation} \label{25)} 
e(\nabla \theta )=E\{ p(t,l)^{2} \} E\{ 2-2\cos \frac{\omega _{0} }{\omega _{s} } \nabla \theta \} 
\end{equation} 
This equation shows if the initial error is 0, the primary noise can be completely canceled at the location of the error microphone. Due to $\mathrm{\nabla }\theta $ is a uniform distributed random variable,  the equation \eqref{25)} can be furtherly derived as 
\begin{equation} \label{26)} 
e(\nabla \theta )=2E\{ p(t,l)^{2} \} (1-\frac{\sin \frac{\omega _{0} \pi }{\omega _{s} } }{\frac{\omega _{0} \pi }{\omega _{s} } } ). 
\end{equation} 

In the equation \eqref{26)}, ${\omega }_s$ is the sampling frequency. The curve of equation \eqref{26)} is plotted in Figure 5. 
\begin{figure}[!t]
    \centering
    \includegraphics[width=9cm]{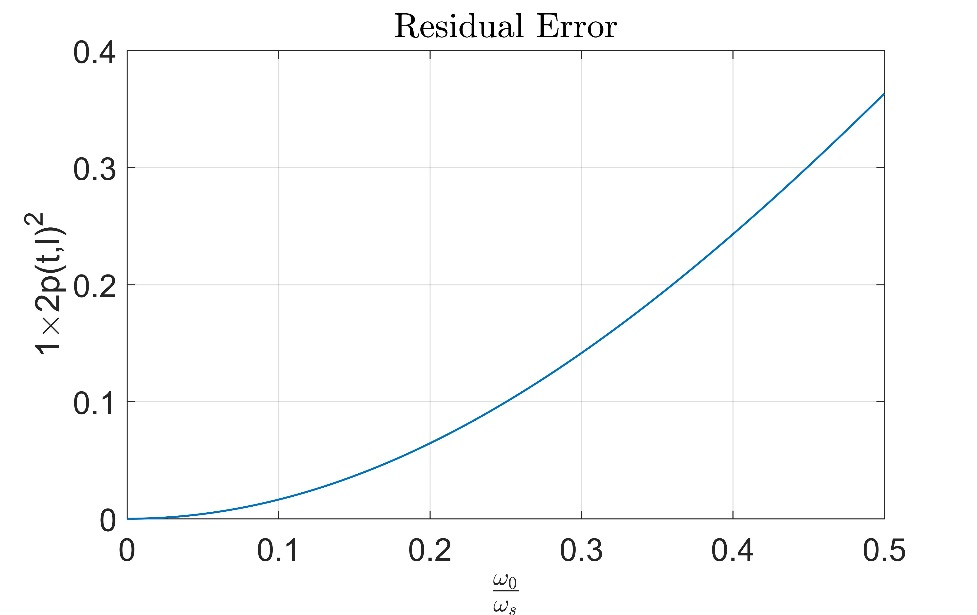}
    \caption{Residual Error vs the frequency of the reference signal}
    \label{fig:5}
\end{figure}

From Figure 5, we can figure out that the residual error will increase with the reference signal frequency compared. In other words, the initial phase error has more considerable influence on tonal noise with higher frequency than low frequency.

\subsection{Broad band primary noise}
To analyze the initial phase noise on a single-channel ANC system when it works on broadband noise, we set the primary noise as a chirp signal (Linear Frequency Modulation) as
\begin{equation} \label{27)} 
p'(t,x)=p_{a} e^{j[\pi mt^{2} -k(t)x]}, 
\end{equation} 
\begin{equation} \label{28)} 
k(t)=\sqrt{\frac{4m^{2} t^{2} +2m}{c_{0}^{2} } }  ,  
\end{equation} 
\begin{equation} \label{29)} 
m=\frac{B}{T_{L} }  .  
\end{equation} 

In the equation, B and $T_L$ are the bandwidth and period of the LFM signal, respectively. Hence, the sound pressure of primary noise at the reference microphone is given by 
\begin{equation} \label{30)} 
p'(t,0)=p_{a} e^{j\pi mt^{2} }. 
\end{equation} 
The reference signal sampled by the error sampling clock should be 
\begin{equation} \label{31)} 
x'''(t)=p_{a} \sum _{n=-\infty }^{\infty }e^{j\pi m(t+\frac{\nabla \theta }{2\pi } T_{s} )^{2} } \delta (t-nT_{s} ).  
\end{equation} 
Furthermore, we assume the optimal control filter just brings in the delay to the input signal and changes the amplitude to negative. Therefore, the anti-noise should be
\begin{equation} \label{32)} 
c'''(t,l)=x(t)*g^{o} (t)*h(t)*s(t)=-p_{a} e^{j\pi m(t+\frac{\nabla \theta }{2\pi } T_{s} -t_{c} -t_{s} )^{2} }.  
\end{equation} 
The disturbance noise at the error microphone should be 
\begin{equation} \label{33)} 
p_{d} (t,l)=p_{a} e^{j\pi m(t-t_{c} -t_{s} )^{2} }.  
\end{equation} 
The square of residual error at the location of the error microphone is derived as  
\begin{equation} \label{34)} 
e^{2} (t)=2p_{a}^{2} \{ 1-\cos [2\pi m(t-t_{c} -t_{s} )\frac{\nabla \theta }{2\pi } T_{s} +\pi m(\frac{\nabla \theta }{2\pi } T_{s} )^{2} ]\}  
\end{equation} 

From the equation, we can know the $\mathrm{\nabla }\theta $ will also cause the worst noise reduction. When the $\mathrm{\nabla }\theta $  is the zeros, the primary noise can be completely canceled. If we substitute the equation \eqref{29)} to \eqref{34)}, we can get
\begin{equation} \label{35)} 
e^{2} (t)=2p_{a}^{2} \{ 1-\cos \{ \frac{B}{T_{L} } \times [2\pi (t-t_{c} -t_{s} )\frac{\nabla \theta }{2\pi } T_{s} +\pi (\frac{\nabla \theta }{2\pi } T_{s} )^{2} ]\} \}.  \end{equation} 
When the $T_L$ is big enough, the $e^2(t)$ is close to 0. From the above two analyses, we know the initial phase error has little affection for the reference signal over a long period.

\subsection{Narrow band primary noise in the Multi-channel ANC system}
\begin{figure}
    \centering
    \includegraphics[width=11cm]{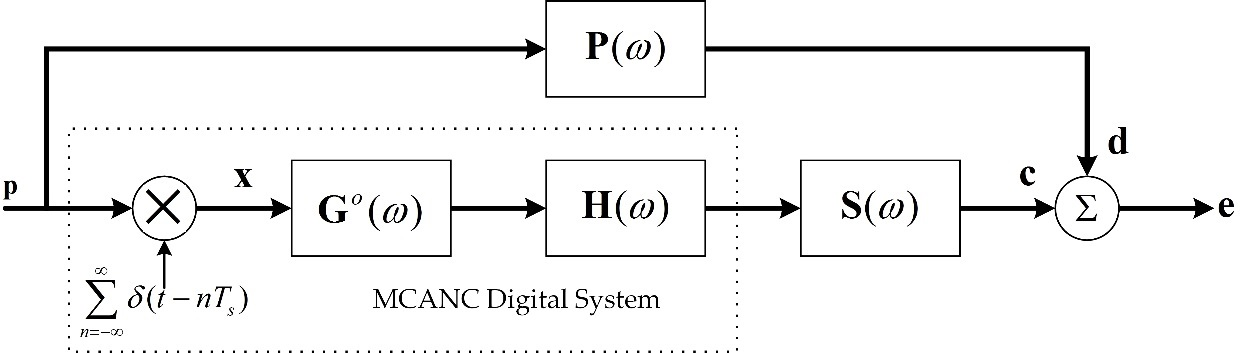}
    \caption{Block diagram of the multichannel ANC system}
    \label{fig:enter-label}
\end{figure}

The block diagram of the multichannel ANC~\cite{shi2023multichannel,luo2022implementation,shi2021block,shi2020multichannel,shi2019practical,shi2017understanding,shi2017multiple}, which works with the optimal control filter, is shown in Figure 6. The primary noise is a sine tone whose frequency is ${\omega }_0$. The sampled reference signal vector can be described as 
\begin{equation} \label{36)} 
\boldsymbol{\mathrm{X}}(\omega )=[X_{0} (\omega ),X_{1} (\omega ),\ldots ,X_{j} (\omega )]^{T}. 
\end{equation} 
When the sampling clock brings in the initial phase error, the sampled reference signal vector is given by 
\begin{equation} \label{37)} 
\boldsymbol{\mathrm{X}}^{'} (\omega )=e^{j\omega _{0} \frac{\nabla \theta }{2\pi } T} [X_{0} (\omega ),X_{1} (\omega ),\ldots ,X_{j} (\omega )]^{T}. 
\end{equation} 
The signal picked up by the error microphone is represented as 
\begin{equation} \label{38)} 
\boldsymbol{\mathrm{e}}(t)=\boldsymbol{\mathrm{d}}(t)-e^{j\omega _{0} \frac{\nabla \theta }{2\pi } T} Fourier^{-1} \{ \boldsymbol{\mathrm{S}}(\omega )\boldsymbol{\mathrm{H}}(\omega )\boldsymbol{\mathrm{G}}(\omega )\boldsymbol{\mathrm{X}}(\omega )\}  
\end{equation} 

We assume the anti-noise can completely cancel the primary noise without the influence of the error sampling clock.  Hence, 
\begin{equation} \label{39)} 
\boldsymbol{\mathrm{d}}(t)-Fourier^{-1} \{ \boldsymbol{\mathrm{S}}(\omega )\boldsymbol{\mathrm{H}}(\omega )\boldsymbol{\mathrm{G}}(\omega )\boldsymbol{\mathrm{X}}(\omega )\} =\boldsymbol{\mathrm{0}}.   
\end{equation} 
By substituting the equation \eqref{39)} into \eqref{38)}, we can get 
\begin{equation} \label{40)} 
\boldsymbol{\mathrm{e}}(t)=\boldsymbol{\mathrm{d}}(t)(1-e^{j\omega _{0} \frac{\nabla \theta }{2\pi } T} ).   
\end{equation}
The exception of the squared equation of \eqref{40)} is obtained from 
\begin{equation} \label{41)} 
E\{ e^{2} (t)\} =E\{ d^{2} (t)\} (1-\frac{\sin \frac{\omega _{0} \pi }{\omega _{s} } }{\frac{\omega _{0} \pi }{\omega _{s} } } ). 
\end{equation}
The outcome is comparable to that of single-channel active noise control (ANC) when the frequency of the reference noise is low, as the initial phase error has a diminished impact on the effectiveness of noise reduction. 


\bibliography{references.bib}\label{refs}
\bibliographystyle{ieeetr}

\end{document}